\documentclass[aps,prl,superscriptaddress,reprint,longbibliography]{revtex4-2}

\usepackage{amsmath}        
\usepackage{amsfonts}
\usepackage{amssymb}
\usepackage{graphicx}

\usepackage{bm}
\usepackage{enumitem}
\usepackage{epsfig}
\usepackage{color}

\usepackage{color}
\usepackage{hyperref}
\usepackage{xcolor,soul}

\begin{document}

\title{Photonic heat transport in three terminal superconducting circuit}
\author{Azat Gubaydullin}
\email[]{azat.gubaydullin@aalto.fi}
\affiliation{QTF Centre of Excellence, Department of Applied Physics, Aalto University, P.O. Box 15100, FI-00076 Aalto, Finland}
\author{George Thomas}
\affiliation{QTF Centre of Excellence, Department of Applied Physics, Aalto University, P.O. Box 15100, FI-00076 Aalto, Finland}
\author{Dmitry S. Golubev}
\affiliation{QTF Centre of Excellence, Department of Applied Physics, Aalto University, P.O. Box 15100, FI-00076 Aalto, Finland}
\author{Dmitrii Lvov}
\affiliation{QTF Centre of Excellence, Department of Applied Physics, Aalto University, P.O. Box 15100, FI-00076 Aalto, Finland}
\author{Joonas T. Peltonen}
\affiliation{QTF Centre of Excellence, Department of Applied Physics, Aalto University, P.O. Box 15100, FI-00076 Aalto, Finland}
\author{Jukka P. Pekola}
\affiliation{QTF Centre of Excellence, Department of Applied Physics, Aalto University, P.O. Box 15100, FI-00076 Aalto, Finland}

\date{\today}
\maketitle

\textbf{
Quantum heat transport devices are currently intensively studied in theory. Experimental realization of quantum heat transport devices is a challenging task. So far, they have been mostly investigated in experiments with ultra-cold atoms and single atomic traps.  Experiments with superconducting qubits have also been carried out and heat transport and heat rectification has been studied in two terminal devices. The structures with three independent terminals offer additional opportunities for realization of heat transistors, heat switches, on-chip masers and even more complicated devices.  Here we report an experimental realization of a three-terminal photonic heat transport device based on a superconducting quantum circuit. Its central element is a flux qubit made of a superconducting loop containing three Josephson junctions, which is connected to three resonators terminated by resistors. By heating one of the resistors and monitoring the temperatures of the other two, we determine photonic heat currents in the system and demonstrate their tunability by magnetic field at the level of 1 aW. We determine system parameters by performing microwave transmission measurements on a separate nominally identical sample and, in this way, demonstrate clear correlation between the level splitting of the qubit and the heat currents flowing through it. Our experiment is an important step in the development of on-chip quantum heat transport devices. On the one hand, such devices are of great interest for fundamental science because they allow one to investigate the effect of quantum interference and entanglement on the transport of heat. On the other hand, they also have great practical importance for the rapidly developing field of quantum computing, in which management of heat generated by qubits is a problem. Therefore, we anticipate that three terminal, and even more complicated, quantum heat transport devices will be increasingly studied in the near future.
}

\section{Introduction}

Recent achievements in superconducting circuit QED techniques in combination with 
ultrasensitive nanoscale thermometry \cite{Blais,Meschke, revmodgiaz} stimulated theoretical discussion and experimental realization of
on-chip refrigerators \cite{Timofeev,tan_quantum-circuit_2017,Partanen,karimi_otto_2016}, quantum heat engines \cite{kosloff}, 
heat rectifiers \cite{GiazottoRect}, interferometers \cite{martinezheat}, and other thermal devices.
Superconducting loops containing 
Josephson junctions and coupled to superconducting resonators are essential parts of such systems because they allow
one to control photonic heat currents by magnetic field. 
Recent examples based on this architecture include quantum heat valve based on the resonator-qubit-resonator 
assembly \cite{QHV} and heat rectifier with unequal frequency resonators \cite{QHR}.
 
\begin{figure}
\includegraphics[width=0.8\columnwidth]{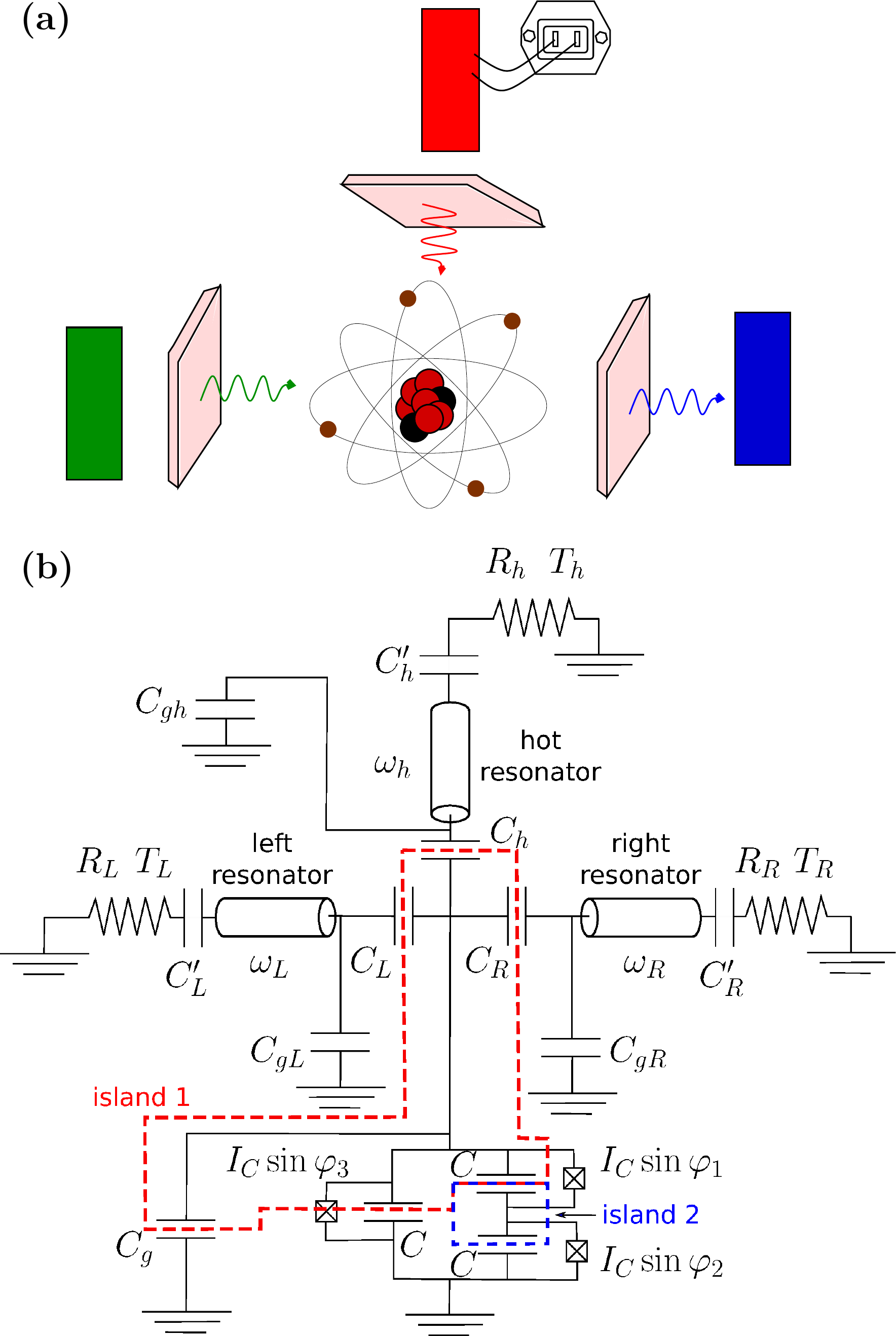}
\caption{(a) Heat transport in a three terminal system containing an artificial atom. 
(b) Schematics of a superconducting circuit, which models our device and realizes the heat transport experiment
sketched in panel (a).}
\label{scheme}
\end{figure} 

So far most experiments in this field have been carried out in two terminal devices. 
However, according to the theory, moving to three terminal setup \cite{Sanchez} should open up new opportunities.
Indeed, there exist theory proposals for thermal transistors \cite{Joulain,Zhang,Majland,KurizkiPRR}, 
heat amplifiers \cite{Liu,Wang}, quantum absorption refrigerators \cite{kosloff2,Segal} 
and thermally pumped masers \cite{Scovil1959,George}, which are heat engines from thermodynamical point of view, in this configuration.
Motivated by these ideas, we have fabricated a three terminal heat transport device containing
an Xmon qubit, which acts as a tunable element and controls photonic heat currents between the
terminals. Basic schematics of our device is shown in Fig. \ref{scheme}. Its main element is
the flux qubit made of a superconducting loop containing three identical Josephson junctions. 
Qubit frequency can be tuned by aplication of magnetic field, which induces the magnetic 
flux inside the loop. The qubit is coupled to three resonators -- the left, the right and the hot one, 
which filter thermal noises emitted by three ohmic resistors. The temperature of the hot resistor can be varied by application of
the heating current to it and the temperatures of the two other resistors are monitored by electronic thermometers.
Heating the hot resistor we bring the whole system into the non-equilibrium steady state and 
vary the heat currents by magnetic flux.   
In this way, we demonstrate the control of the photonic heat power at the level
of $10^{-18}$ W. In order to get more information about the system paramters, 
we have performed microwave transmission measurements on the nominally identical twin sample.
We have also developed a theory model, which reasonably well explains the experimental findings.
We are confident that further technological developments will soon permit
practical implementation of the interesting theoretical proposals mentioned above and
investigation of the effects of quantum coherence on the performance of heat transport devices \cite{Segal}.

%The paper is organized as follows: in Sec. \ref{experiment} we describe the sample design,
%in Sec. \ref{Theory} --- the theoretical model, in Sec. \ref{Results} we present and discuss
%the experimental results and in Sec. \ref{Summary} we summarize the discussion.

\section{Results}
\subsection{Experimental}
\label{experiment}

\begin{figure}
\includegraphics[width = 1.0\columnwidth]{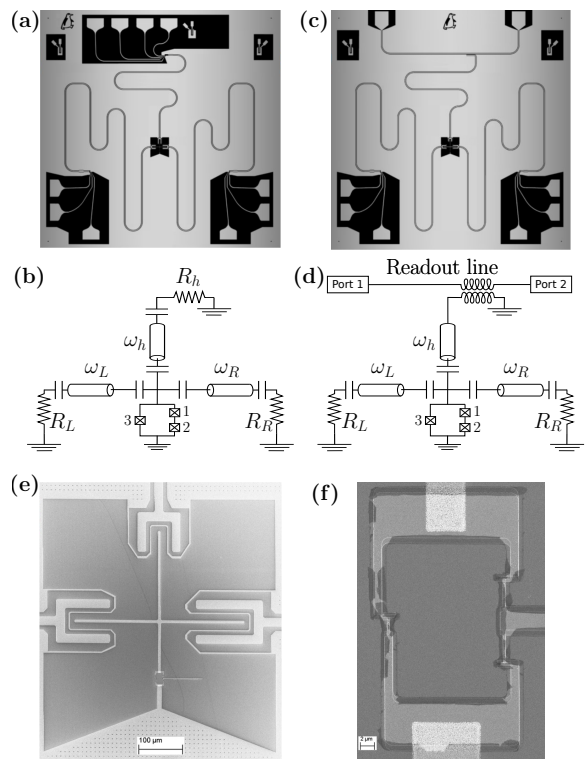}
\caption{ Images of the sample I used in the heat transport experiment (a) 
and of the sample II used for the microwave spectroscopy (c).
Both samples contain a cross-shaped Xmon-type superconducting qubit, 
which is capacitively coupled to three superconducting coplanar waveguide resonators. 
(b) Equivalent lumped-element circuit of the sample I and (d) of the sample II. 
(e) Scanning electron micrograph of the Xmon qubit island (scale bar has the length 2 $\mu$m,)
and (f) of the flux qubit loop with three Josephson junctions.}
\label{device}
\end{figure}

In order to fully characterize the system, we have fabricated two samples with
nominally identical parameters on the same wafer, but designed for different measurement setups. 
The parameters of these samples may differ due to fabrication uncertainities, which we roughly estimate as 10 \%.
The first of these devices, sample I shown in Fig. \ref{device}a, contains three resistors and was used for 
DC measurements of the photonic heat currents between the resistors. 
In the second device, sample II presented in Fig. \ref{device}c, 
we have replaced one of the resistors by a transmission line, which allowed us to carry out
detailed spectroscopic measurements of the qubit in the microwave frequency range.

\begin{figure}
\includegraphics[width = 1.0\columnwidth]{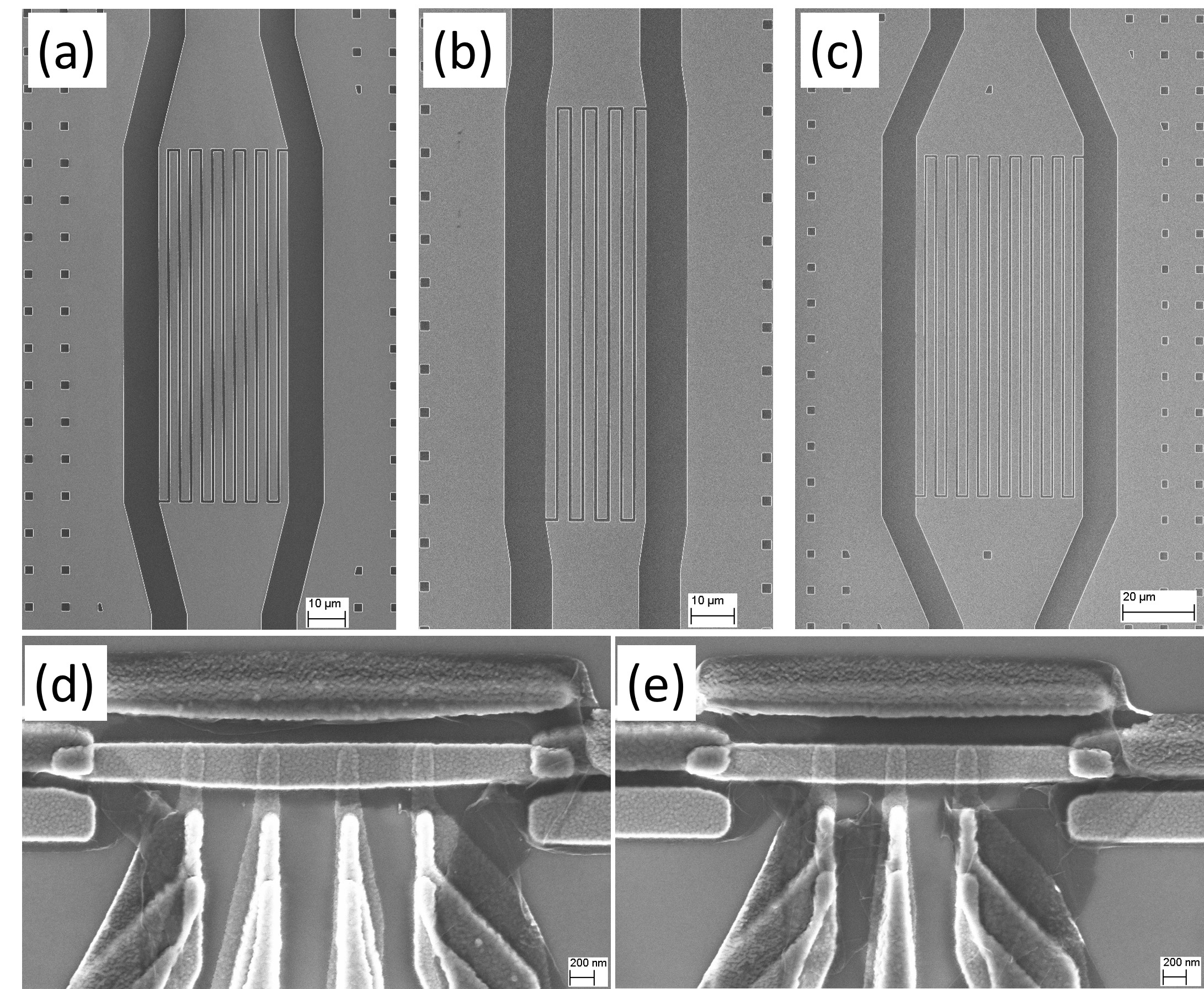}
\caption{\label{CapRes}(a) Scanning electron micrograph of the capacitors between resonators and right resistor 
$C'_R=48 fF$ (c),  hot resistor  $C'_{h}=30 fF$ (b), and left resistor $C'_{L}=70 fF$ (c). 
The indicated capacitance values are obtained from COMSOL simulations. 
(d) hot resistor $R_{h}=5$ $\mathrm{\Omega}$ (d).  
(e) left and right resistors have the same design and nominal values
$R_{L}=R_R=4$ $\mathrm{\Omega}$ (e).}
\end{figure}

Sample I contains centrally located cross-shaped superconducting island, which is capacitively coupled to three superconducting $\lambda/2$ 
coplanar waveguide (CPW) resonators with the characteristic impedance $Z_0=50$ $\Omega$ as 
schematically shown in Fig.~\ref{device}b. 
The bottom leg of the cross is connected to the qubit, see Fig.~\ref{device}e, 
which is realized as a superconducting loop with three asymmetrically arranged Josephson junctions, 
similar to the conventional persistent-current flux qubit \cite{Orlando}. 
The SEM image of the three junction loop is presented in Fig.~\ref{device}f.
The top resistor, $R_h$, is heated by DC bias current
and serves as the heater for the whole system. 
Two other resistors, $R_L$ and $R_R$, are passively
heated by the power emitted by the hot resistor and their temperatures are monitored.
The resistors are realized as copper islands with the resistances $R_L=R_R=4$ $\Omega$
and $R_h=5$ $\Omega$ and the volumes ${\cal V}_L={\cal V}_R=0.036$ $\mu$m$^3$ and ${\cal V}_h=0.048$ $\mu$m$^3$. 
Their images are shown in Figs. \ref{CapRes}d,e.
The coupling between the qubit and the resonators is mediated by the capacitors 
with the designed values
$C_{L}=3.81$ fF, $C_{h}=4.51$ fF, and $C_{R}=4.66$ fF. 
Additional capacitors having designed values
$C'_L=70$ fF, $C'_h=30$ fF and $C'_R=48$ fF are inserted between the resonators and the resistors
in order to keep the quality factors of the resonators sufficiently high. 
Their images are shown in Fig. ~\ref{CapRes}a,b and c. The quality factors of the resonators are given as
$Q_i=\pi/(2 \omega_i^2 Z_0 R_i {C_i'}^2)$ for $i=$ L, R or h \cite{Esteve2008}, where $\omega_i=2\pi f_i$ are the resonator frequencies.
With the parameters listed above and with measured resonator frequencies %  reported in Sec. \ref{spectroscopy}
we estimate $Q_L=2450$, $Q_R=4690$ and $Q_h=2720$.
Finally, the capacitances to the ground, schematically shown in Fig. \ref{device}b,
are found to be $C_{gL}=48.8$ fF, $C_{gR}=54$ fF, $C_{gh}=60.4$ fF and $C_g=42$ fF.
All capacitance values listed above, as well as
the self-capacitance of the Xmon island \cite{Barends}  
and the design of the SQUID are optimized by device simulation using finite element 
modeling COMSOL Multiphysics software packet. 
The designed value for the charging energy is   
$E_C/h =0.3$  $\mathrm{GHz}$, and for the Josephson energy - $E_{J}/h=4.6$  GHz.
The latter value corresponds to the critical current of a single junction $I_C=9.3$ nA.  
The resonators are designed in such a way that the frequencies of the left and the right
resonators are close to each other and to the qubit frequency, while the frequency of the hot resonator is approximately
two times higher,  see the \textit{spectroscopy} section for details.

As shown in Figs. \ref{CapRes}d,e,
each normal metal (N) resistor has aluminium superconducting (S) probes, separated by a thin insulating (I) layer, 
which we use for thermal control and readout.  
Four superconducting probes allow us to control and simultaneously  measure the resistor temperature. 
We change the temperature by applying voltage bias between the two superconducting probes. 
For bias voltages above the superconducting energy gap the resistor is heated up, while for voltages below the gap it is cooled. 
The electronic temperature readout is performed by applying current bias between another pair of NIS tunnel junctions 
in a SINIS configuration \cite{revmodgiaz}. The details of electronic thermometry are presented in the \textit{methods} section. 
The Andreev mirrors at the aluminium-copper boundaries on both sides of each resistor help to localize the heat,
and, at the same time, they ensure good electric contact between the resistors, the ground electrode and the resonators.
The dominating heat relaxation channel in the resistors is the electron-phonon coupling. 
Power leakage to the phonons is estimated in the usual way: 
\begin{eqnarray}
P_\mathrm{el-ph} = \Sigma \mathcal{V} (T_\mathrm{el}^5 - T_\mathrm{ph}^5), 
\label{Peph}
\end{eqnarray}
where $\Sigma=2\times 10^{9}$ Wm$^{-3}$K$^{-5}$ is the electron-phonon coupling constant, 
$\cal{V}$ is volume of the copper resistor, $T_\mathrm{el}$ 
and $T_\mathrm{ph}$ are the temperatures of electrons and phonons respectively. 
In the steady state the power dissipated in the resistor equals to the power leaking to the phonons.
Thus, measuring the electronic temperature $T_\mathrm{el}$ with the thermometer and knowing the substrate temperature  $T_\mathrm{ph}$
one can easily estimate the dissipated power from Eq. (\ref{Peph}). 
Alternatively, knowing Joule heating power dissipated in the hot resistor, 
from Eq. (\ref{Peph}) one can estimate its electronic temperature $T_\mathrm{el}$.

Sample II, shown in Figs. \ref{device}c,d, and designed for the microwave spectroscopy, has nominally the same design and parameters.
However, this sample does not have a hot resistor and the top resonator serves for diagnostic purposes.
The latter has the same length as the top resonator of the sample I, but it is realized as $\lambda/4$-resonator with its upper end grounded.
As a result, its frequency becomes two times lower than that in the sample I.  
This modification in the design brings the diagnostic resonator frequency closer
to the frequencies of the two other resonators and to the transition frequency between the two lowest levels of the qubit,
and makes the spectroscopy more accurate.

\subsection{Theory model}
\label{Theory}

We describe the flux qubit loop containing three identical Josephson junctions with the theory model of Ref. \cite{Orlando}.
The loop contains two superconducting islands.
The first island, denoted as I1, is large and includes the cross-like aluminium electrode. 
It is restricted by coupling capacitors 
$C_L$, $C_h$ and $C_R$ and the Josephson juctions numbered 1 and 3, 
as shown in Fig. \ref{scheme}b by red dashed line. 
The second island I2 is much smaller, in Fig. \ref{scheme}b it is indicated by the blue dashed line 
and sandwiched between the junctions 1 and 2. We introduce the operators $n_{I1}$ and $n_{I2}$ representing
the number of Cooper pairs in the islands I1 and I2, 
and the operators $\varphi_i$ with $i=1,2$ or 3, corresponding to the Josephson phase differences across the $i$th junction. 
The Hamiltonian of the flux qubit is \cite{Orlando}
\begin{eqnarray}
 && H=4\sum_{k,l}n_k n_l  (E_C)_{kl}
\nonumber\\
 && -\, E_J \left[\cos\varphi_2+ \cos\varphi_3+\cos(\varphi_{\rm ext}+\varphi_2-\varphi_3)-3\right],
 \label{ham}
\end{eqnarray}
where $\varphi_{\rm ext}=2\pi\Phi/\Phi_0$, $\Phi$ is the normalized magnetic flux threading the loop,
$\Phi_0 = h / 2e$ is the magnetic flux quantum,  $E_J=\hbar I_C/2e$ is the Josephson energy of a single junction having 
the critical current $I_C$, 
$(E_C)_{kl}=e^2(C^{-1})_{kl}/2$, and $(C^{-1})_{kl}$ are the elements of the inverse of the $2\times 2$ capacitance matrix 
\begin{equation}
 C=\left[ {\begin{array}{cc}
  C_{I1}&-C\\
  -C&C_{I2}\\
     \end{array} 
  } \right].
\label{rates_matrix}
\end{equation}
Here $C_{I1}=C_L+C_h+C_R+C_g+2C$ and $C_{I2}=2C$ are the total capacitances of the islands 1 and 2, respectively. 
We diagonalize the Hamiltonian \ref{ham} numerically in the basis of 
two dimensional plane waves 
having the form  $\exp(-i n_{I1} \varphi_3-i n_{I2} \varphi_2)/2 \pi$.
Since $E_J\gg E_C=e^2/2C_{I1}$ for all values of $\Phi$ except the narrow region close to $\Phi_0/2$, we neglect
weak dependence of the eigen-energies $E_n$ of the Hamiltonian (\ref{ham}) on the gate charges induced by,
for example, charged impurities.

In order to fit the experimental data on the heat power transmitted between the resistors we use  
the Landauer formula for the total power $P_i$ carried by photons and dissipated in the resistor with the number $i$,
\begin{eqnarray}
P_i^{\rm ph} &=& \sum_{j\not=i} P_{ij}^{\rm ph},
\label{Pi}\\
P_{ij}^{\rm ph} &=& \int_0^\infty \frac{d\omega}{2\pi}\tau_{ij}(\omega)
\left[ \frac{\hbar\omega}{e^{\frac{\hbar\omega}{k_BT_j}}-1} - \frac{\hbar\omega}{e^{\frac{\hbar\omega}{k_BT_i}}-1} \right]. 
\label{Pij}
\end{eqnarray}
Here the indexes $i$ and $j$ enumerate the resistors, i.e. they can take the values $L,R$ or $h$,
and $T_i$ are the resistor temperatures. Here we have also introduced the heat currents $P_{ij}^{\rm ph}$ flowing from the resistor $j$
to the resistor $i$.
In order to derive the expression for the
photon transmission probabilities $\tau_{ij}(\omega)$, we linearize Josephson dynamics and
replace all three junctions in the loop by the identical inductors with the inductance $L=\hbar/2eI_C$.
Solving the corresponding Kirchhoff equations with thermal Nyquist noise sources connected in parallel with 
the resistors, as outlined in Ref. \cite{Pascal} for example, we find
\begin{eqnarray}
\tau_{ij}(\omega) = \frac{{\rm Re}\left[\frac{1}{Z_i(\omega)}\right]{\rm Re}\left[\frac{1}{Z_j(\omega)}\right]}
{\left|\frac{1}{Z_J(\omega)}+\frac{1}{Z_L(\omega)}+\frac{1}{Z_R(\omega)}+\frac{1}{Z_h(\omega)}\right|^2}.
\label{tau}
\end{eqnarray}
Here we have introduced the impedances of the three segments of the electric circuit depicted in Fig. \ref{scheme}b,
which contain individual resonators, resistors and coupling capacitors. They are defined as 
\begin{eqnarray}
Z_j(\omega) &=& \frac{1}{-i\omega C_j} + \frac{1}{-i\omega C_{gj} + Z_{rj}^{-1}(\omega)},
\end{eqnarray}
where $Z_{rj}(\omega)$ have the form
\begin{eqnarray}
Z_{rj}(\omega)=Z_0\frac{\left(R_j+\frac{1}{-i\omega C'_j}\right)\cos\omega t_j - iZ_0\sin\omega t_j}
{- iZ_0\cos\omega t_j - i\left(R_j+\frac{1}{-i\omega C'_j}\right)\sin\omega t_j }.
\end{eqnarray} 
Here  $t_j$ is the
flight time of a photon between the two ends of a given resonator which is proportional to its
length. The impedance of the three junction loop for the flux values $|\Phi|\leq \Phi_0/2$ takes the form
\begin{eqnarray}
\frac{1}{Z_J(\omega)} = -i\omega \left(C_g+\frac{3}{2}C\right) + \frac{3eI_C}{-i\hbar\omega}\cos\left( \frac{2\pi}{3}\frac{\Phi}{\Phi_0} \right), 
\end{eqnarray} 
and it should be periodically extended with the period $\Phi_0$ for $|\Phi|>\Phi_0/2$.
Transmission probabilities (\ref{tau}) exhibit multiple peaks centered at frequencies 
corresponding to the modes of the resonantors and one additonal narrow peak at the resonance frequency
of the three junction loop 
\begin{eqnarray}
\omega_0(\Phi) = \sqrt{\frac{3eI_C}{\hbar C_\Sigma}\cos\left( \frac{2\pi}{3}\frac{\Phi}{\Phi_0} \right)},
\;\; |\Phi|<\frac{\Phi_0}{2}.
\label{omega_q}
\end{eqnarray}
Here $C_\Sigma = C_g + 3C/2 + C_L + C_R + C_h$ is effective capacitance of the qubit, which is similar, but slightly
different from $C_{I1}$. 
The frequency $\omega_0$ is close to the exact transition frequency between the two lowest levels of 
the non-linear qubit $2\pi f_{01}=(E_1-E_0)/\hbar$,
but deviates from it in the vicinity of the flux point $\Phi=\Phi_0/2$, where qubit anharmonicity becomes 
significant. 

As shown in Ref. \cite{Pascal},
the Landauer formula (\ref{Pi},\ref{Pij}) can be derived from the Kirchhoff's equations relating 
the Fourier components $I_{i,\omega}$ of the four fluctuating in time currents $I_i(t)$, which flow
from the central cross shaped island of the device in the left ($i=L$), the hot ($i=h$), the right ($i=R$) resonators and in the qubit ($i=q$), 
with the island potential $V_{isl}$,
\begin{eqnarray}
I_{i,\omega}=\frac{V_{isl,\omega}}{Z_i(\omega)}+\xi_{i\omega}.
\label{Ij}
\end{eqnarray} 
Here $\xi_{i,\omega}$ are the Fourier components of the noise currents with the spectral densities determined by the 
fluctuation-dissipation theorem,
\begin{eqnarray}
\langle|\xi_{i,\omega}|^2\rangle  = \,{\rm Re}\left[\frac{1}{Z_i(\omega)}\right]\hbar\omega\coth\frac{\hbar\omega}{2k_BT_i}.
\label{noise}
\end{eqnarray}
These noises are generated by the resistors $R_i$ and acquire the spectrum (\ref{noise}) close to the island
due to the filtering effect of the resonators.
The qubit does not generate any noise, which means $\xi_{q,\omega}=0$. 
Eqs. (\ref{Ij}) should be supplemented by the current conservation condition
\begin{eqnarray}
\sum_j I_{j,\omega} = 0.
\label{conserv}
\end{eqnarray}
The power dissipated in the resistor $R_i$ is expressed as
\begin{eqnarray}
P_i = \langle I_i(t) V_{isl}(t) \rangle = \int\frac{d\omega}{2\pi} \langle I_{i,\omega} V_{isl,\omega}^* \rangle.
\label{Pii}
\end{eqnarray}
Solving Eqs. (\ref{Ij},\ref{conserv}) we express the currents $I_{j,\omega}$ and the potential $V_{isl,\omega}$ via the three noises
$\xi_L,\xi_h$ and $\xi_R$. Substituting the result in Eq. (\ref{Pii}) and taking the avarages with the aid of Eq. (\ref{noise}), we arrive
at Eqs. (\ref{Pi},\ref{Pij}).

As we mentioned, Eqs. (\ref{Pij},\ref{tau}) rely on replacing the non-linear Josephson junctions by linear inductors.
This approximation is formally valid in the two limits: at low temperatures $k_BT_i\ll U_b$ and at high temperatures $k_BT_i\gg U_b$,
where $U_b$ is the height of the potential barrier in the potential (\ref{ham}) varying from
$U_b=4E_J$ at $\Phi=0$ to $U_b=E_J/2$ at $\Phi=0.5\Phi_0$. In the former case, Josephson phases fluctuate
close to the bottom of the potential well where one can use harmonic approximation; in the latter case one can put $E_J=0$, which
again makes the system linear and the Landauer formula (\ref{Pij},\ref{tau}) valid. In the intermediate regime $k_BT_i\sim U_b$
the non-linearity of the junctions is important, but even in this case Eqs. (\ref{Pij},\ref{tau}) reasonably well 
describe the heat transport \cite{Thomas}.

%\subsection{Results and discussion}
%\label{Results}
%In this section we present the experimental results and compare them
%with the theory. In Sec. \ref{spectroscopy} we discuss microwave spectroscopy
%and in Sec. \ref{heat} --- heat transport measurements.

\subsection{Microwave spectroscopy}
\label{spectroscopy}

\begin{figure}
\includegraphics[width = \columnwidth]{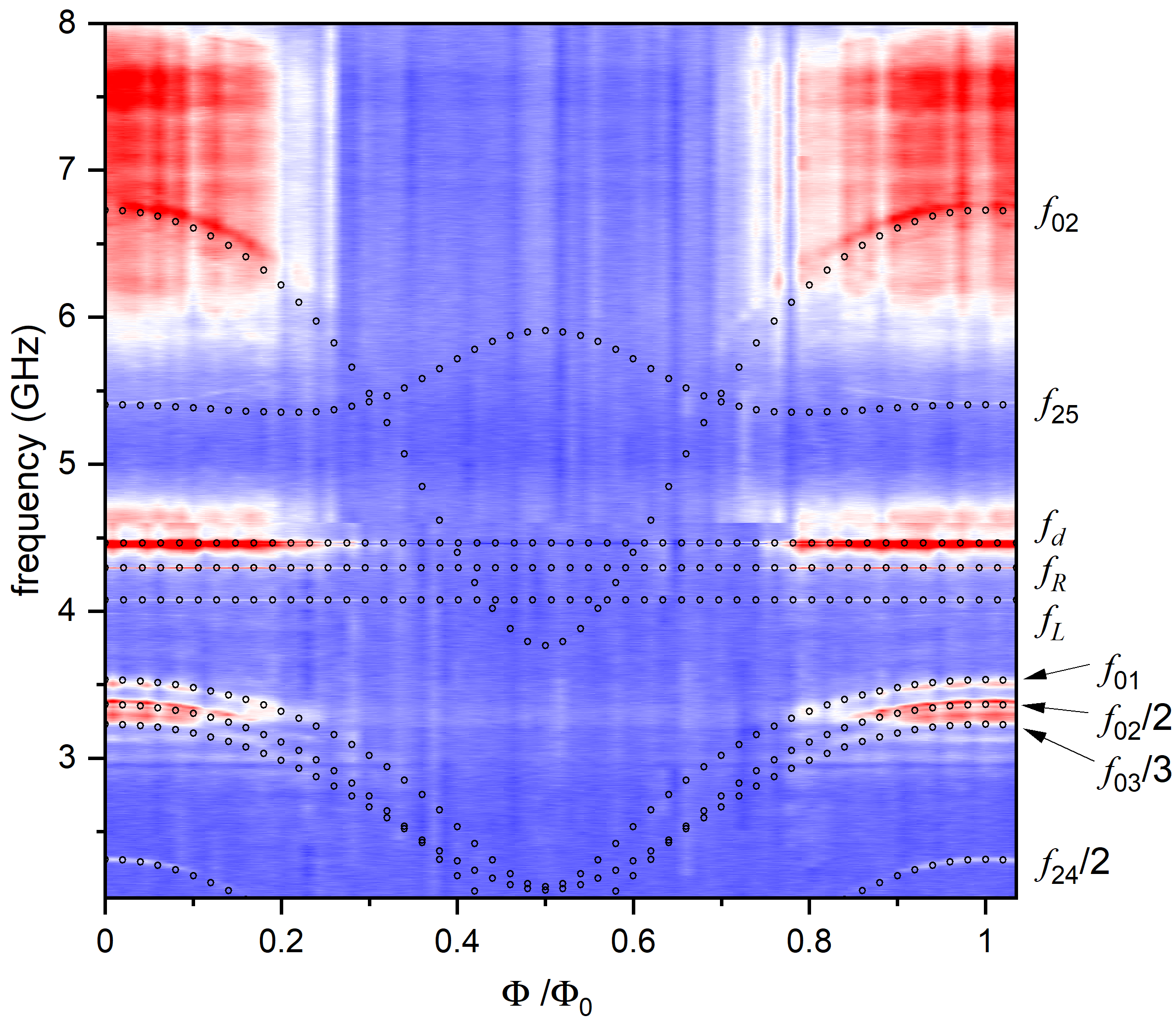}
\caption{ Results of the two-tone spectroscopy: color plot of the transmission coefficient $|S_{21}|$
as a function of the normalized magnetic flux $\Phi/\Phi_0$ and of the frequency $f$.
Cross symbols indicate the modes of the resonators and the dotted lines are
the theory predictions for the single and two photon interlevel transitions. }
\label{spectr}
\end{figure}

\begin{figure}
\includegraphics[width = 0.8\columnwidth]{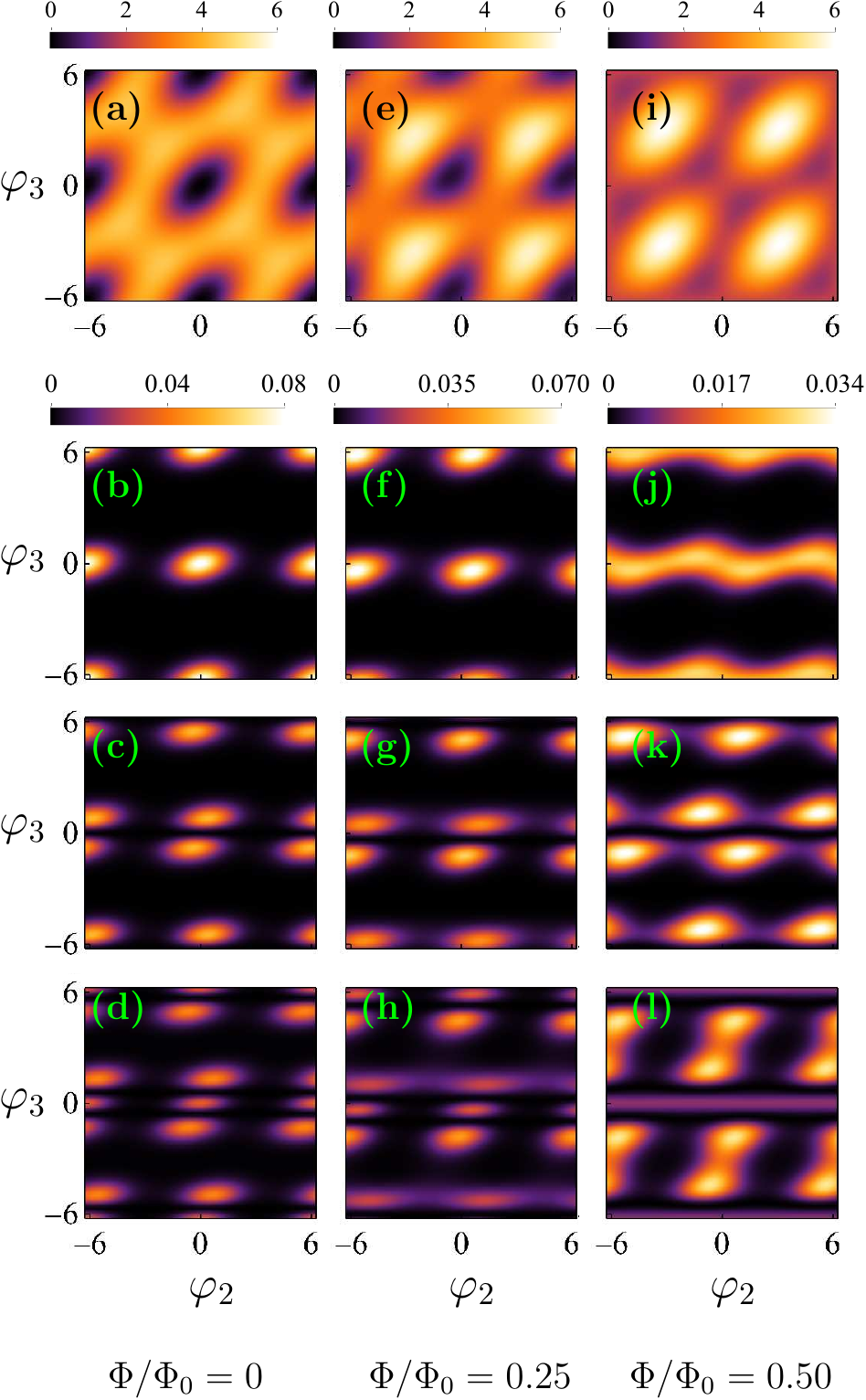}
\caption{ Two dimensional potential of the flux qubit for three 
different values of the magnetic flux: $\Phi=0$ (a), $\Phi=0.25\Phi_0$ (e) and $\Phi=0.5\Phi_0$ (i).  
The squared wave functions of the ground state with $n=0$ are shown in panels (b), (f) and (j);
of the first excited level with $n=1$ -- in panels (c), (g) and (k);
and the second excited level with $n=2$ -- in panels (d), (h) and (l).}
\label{wave_fn}
\end{figure}

In order to obtain accurate information about the device paramters, 
we performed microwave spectroscopy on the Sample II.
In Fig.~\ref{spectr} we show the results of the two-tone spectroscopy
and plot the absolute value of the transmission coefficient $|S_{21}|$ between the ports 2 and 1, shown 
in Fig. \ref{device}d,
as a function of the magnetic flux $\Phi$ and the frequency of the probe signal. 
The spectroscopy reveals a series of lines. 
In particular, we observe single photon transitions at frequencies $f_{01}$, $f_{12}$, $f_{02}$ and $f_{25}$, 
where $f_{ij}=(E_j-E_i)/h$ is the transition frequency between the energy levels of the qubit $E_i$ and $E_j$, and several 
two-photon transitions  corresponding to the frequencies $f_{02}/2$, $f_{24}/2$.  

\begin{figure*}
\includegraphics[width=\textwidth]{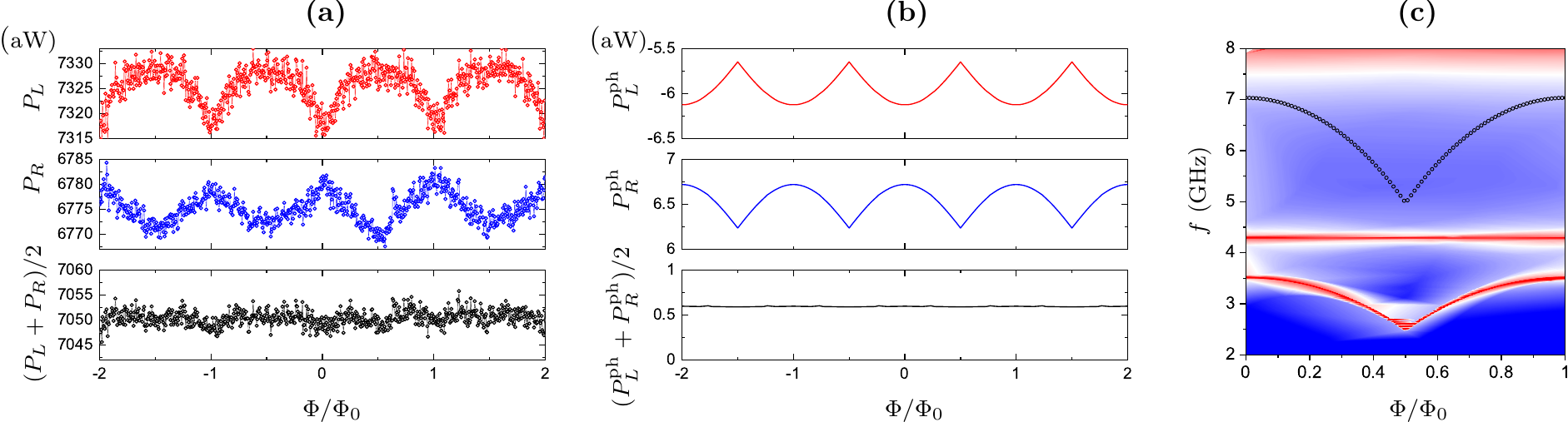}
\caption{(a) Experimentally observed flux dependent 
heat powers dissipated in the left resistor ($P_L$, red dots), in the right resistor ($P_R$, blue dots) and
the average value $(P_L+P_R)/2$ (black dots). 
Heating voltage $V_h=0.9$ mV has been applied to the hot resistor, which resulted in
the resistor temperatures $T_h=393$ mK, $T_L=168$ mK and  $T_R=164$ mK. The temperature of the mixing chamber was $T_{\rm MXC}=127$ mK. 
(b) Theory prediction, based on Eqs. (\ref{Pi},\ref{Pij}), for the photonic heat powers dissipated in the left ($P_L^{\rm ph}$ red line) 
and the right ($P_R^{\rm ph}$ blue line) resistors. Black line in the bottom panel is the average value $(P_L^{\rm ph}+P_R^{\rm ph})/2$.
We assumed the following resistor temperatures: $T_h=393$ mK, $T_L=176$ mK and  $T_R=156$ mK.
(c) The sum of the photon transmission probabilities $\tau_{Lh}(\omega,\Phi)+\tau_{Rh}(\omega,\Phi)$, given by Eq. (\ref{tau}),
assuming the same fit parameters as in the panel (b). Black circles indicate the double frequency of the three junction loop, $2\omega_0(\Phi)/(2 \pi)$,
where $\omega_0(\Phi)$ is defined in Eq. (\ref{omega_q}). 
}
\label{Power}
\end{figure*}

Next, we compare the experimental results with the theory model based on the Hamiltonian (\ref{ham}).
Adjusting the model parameters, we have managed to fit the positions of the experimental spectral lines with high accuracy. 
In this way, we have obtained the total capacitance of the Xmon island $C_{I1}=70.26$ fF and 
the junction capacitances $C=C_{I2}/2=4.13$ fF.
From the same fit we have also obtained the values of the Jospehson energy of a single junction, $E_J/h=4.86$ GHz 
and  the charging energy of the island I1, $E_C/h=276$ MHz.
The errors in the fit is estimated as 3\% for $E_J$ and  4\% for capacitance $C_{I1}-2 C_J$.
All these values are close to the designed ones.
In addition, microwave spectroscopy provides accurate values of the resonator frequencies,
$f_L=4.074$ GHz, $f_R=4.292$ GHz%({\bf are these values correct?})
, and the frequency of the diagnostic
$\lambda/4$-resonator $f_d=4.464$ GHz with its quality factor $Q_d=935$. 
The resonator lines shown in Fig.~\ref{spectr} have
small dispersive shifts induced by the coupling to the qubit, however, they are not visible on
the scale of the figure. For example, for the right resonator we obtained the dispersive shift $\chi_R=1.8$ MHz. 
Importantly, the qubit frequency $f_{01}(\Phi)$ does not cross the frequencies of the resonators
and stays below them for all values of the magnetic flux. In this regime the coupling constants between the resonators
and the qubit can be estimated from the dispersive shifts as $g_{j}=\sqrt{(f_{j}-f_{01}^{\max})\chi_j}$, where $j=L,R,d$ and
$f_{01}^{\max}=3.6$ GHz is the maximum $0 \leftrightarrow 1$ transition frequency. In this way we obtain
$g_R=35$ MHz. In the next section 
we use these parameters to describe the heat transport in the Sample I, which has been fabricated
is the same way. However, sample to sample scattering of the capacitances and the junction resistances during the fabrication 
can reach 10\%, therefore we slightly adjust the capacitance $C_{gR}$ to achieve better fit.

Having determined the system parameters from the fits, 
in Fig. \ref{wave_fn} we plot the two-dimensional potential  of the three junction loop, defined in Eq. \ref{ham},
for three values of the magnetic flux, $\Phi/\Phi_0=0,$ $0.25$ and $0.5$.
In the same figure we also plot the squared absolute values of the wavefunctions of the three lowest energy levels. 
We observe that at flux values $\Phi=0$ and $\Phi=0.25\Phi_0$ 
the wavefunction of the $n$th level shows $n$ nodes in the $\varphi_2$ direction, as expected for a one-dimensional potential.
Thus, in this case the two-dimensional wave function can be approximately factorized into the product $\Psi_n(\varphi_2,\varphi_3)\approx \psi_n(\varphi_2)\psi_0(\varphi_3)$, where $\psi_0(\varphi_3)$ is the ground state
wave function in $\varphi_3$ direction. Since such factorization is exact for a harmonic potential well, we conclude that
at these flux values the linearized model of the qubit, on which Eqs. (\ref{Pij},\ref{tau})
are based, should work reasonably well.   
However, at $\Phi=0.5\Phi_0$  this approximation is no longer accurate because  
the potential well becomes shallow and hence anharmonic.

\subsection{Photonic heat transport}
\label{heat}

\begin{figure*}
\includegraphics[width = 1.5\columnwidth]{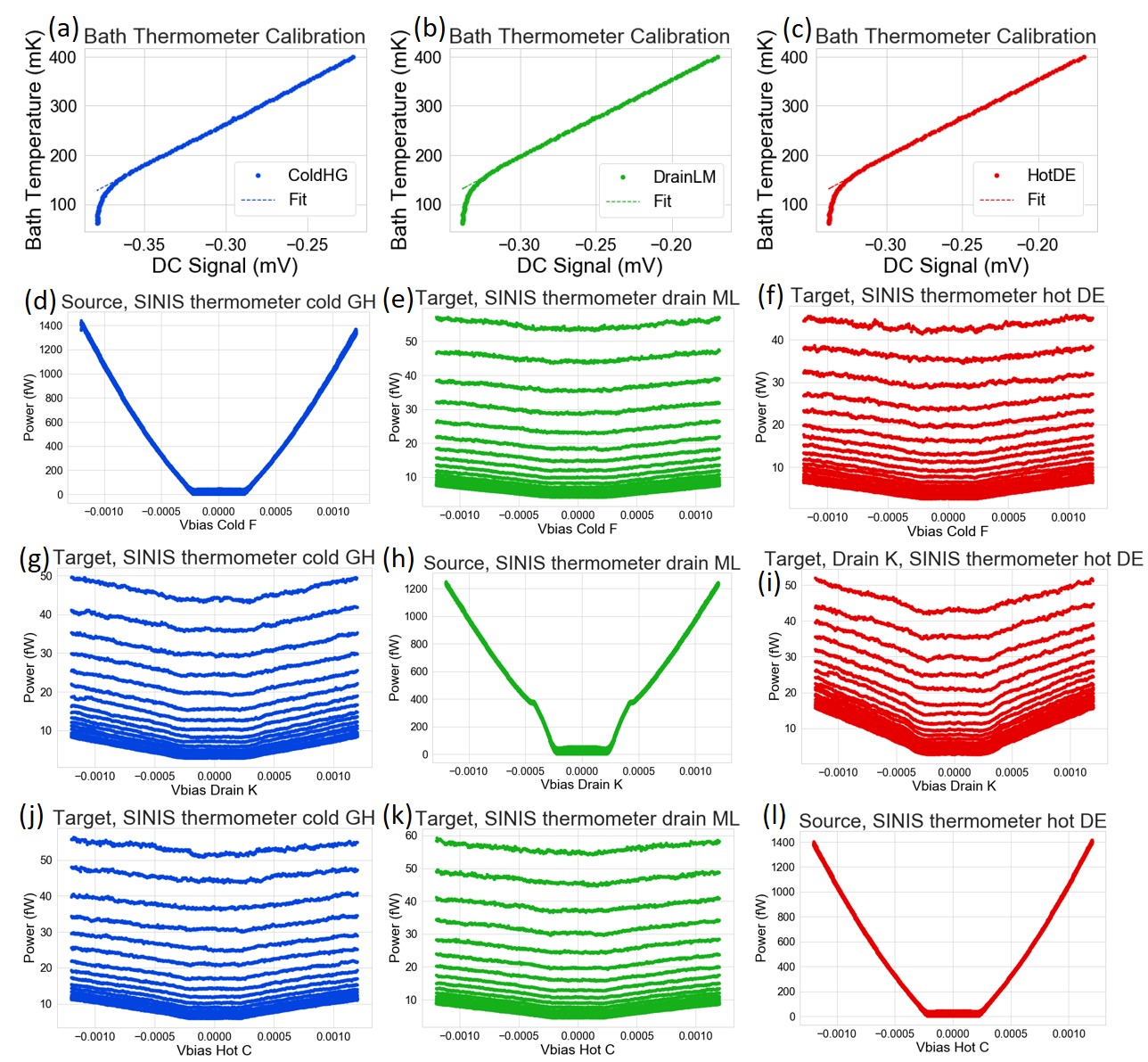}
\caption{\label{Calibration} Calibration of the SINIS thermometer connected to the left resistor (a), to the right resistor (b)
and to the hot resistor (c). In all thse plots the horizontal axis shows the voltage drop across the SINIS structure 
at fixed bias current and the vertical axis -- the temperature of the mixing chamber.  
Panels (d), (h) and (l) show the dependence of the power dissipated in, respectively,  left, right and hot resistors  versus
heating voltage applied to the same resistor. Panels (g) and (j) show the non-local response in the system, where 
the power dissipated in the left resistor is plotted versus the heating voltages applied to the right (g) and to the hot (j) resistors.
We also plot the non-local response of the right resistor on the bias voltages applied to left (e) and to the hot ((k) resistors,
and of the hot reistor on the bias applied to left (f) and right (i) resistors.}
\end{figure*}

We performed heat transport measurements in  Sample I, which has nominally identical parameters with  Sample II. 
The hot resistor was heated up by the DC voltage $V_h$, and the temperatures of the left and the right resistors were monitored with
SINIS thermometers, as it was explained above. We will focus on the data taken at the heating voltage $V_h=0.9$ mV. 
First, we have estimated the power dissipated in the hot resistor as $P_h=I_hV_h/2$ (see the calibration curve in Fig. \ref{Calibration}l), 
and further using Eq. (\ref{Peph}) we have found the temperature of the hot resistor,  $T_h=393$ mK.
This temperature is rather high and, therefore, cannot be accurately measured by the thermometer.
The  temperatures of the left and of the right resistors, which were lower, have been directly measured by SINIS thermometers 
and have been found to be $T_L=168$ mK and  $T_R=164$ mK.
Varying the magnetic flux applied to the SQUID loop, we have observed small oscillations of these temperatures 
with the amplitudes  $\delta T_{L,R}\sim 0.01$ mK. 
The measured temperatures have been converted to powers dissipated in the left ($P_L$) 
and in the right ($P_R$) resistors using Eq. (\ref{Peph}).
They are plotted in Fig.~\ref{Power}a as functions of the magnetic flux $\Phi$. 
$P_L(\Phi)$ and $P_R(\Phi)$ oscillate with the period $\Phi_0$ following the qubit frequency $f_{01}(\Phi)$ shown in Fig. \ref{spectr}.
These oscillations have opposite signs for the two resistors, i.e when $P_L$ increases  $P_R$ goes down and vice versa.
The oscillation amplitudes are similar for both resistors, that is why the average value $(P_L+P_R)/2$, shown by black dots,
is almost constant. This interesting observation suggests that the flux dependent contributions to $P_L$ and $P_R$ originate from
the heat current between these two resistors $P_{LR}$.
Similar behavior of the dissipated powers $P_L,P_R$ has been observed for other values of the heating voltage $V_h$.     

In order to understand the flux dependence of the powers $P_L$ and $P_R$ better, we
have numerically evaluated the photonic heat currents (\ref{Pi},\ref{Pij}) for a circuit depicted in Fig. \ref{device}b. 
Theoretical results are very sensitive to the parameter values. For example, the variation of the capacitances $C_L$ and $C_R$ 
within 10\%, which is the estimated fabrication error, may change the heat fluxes by the factor $10^3 - 10^4$.
The origin of such sensitivity is simple --- any mismatch between the narrow spectral lines of the resonators and of the qubit strongly reduces
the heat flow. For this reason, we do not aim at the perfect fit, our goal is to show that 
with the nominal parameters of the experiment the theoretical model produces qualitatively similar results.
Thus, for this simulation we have 
used the values of the resistances and capacitances given in sections \textit{experiment} and \textit{spectroscopy}.
We have made only few adjustments of the parameters in order to increase the oscillation amplitudes of the heat powers $P_L,P_R$. 
Namely, we have chosen slightly larger value for the ground capacitance $C_{gR}=70.5$ fF and thus
brought the right resonator in resonance with the left one, so that $f_L=f_R=4.292$ GHz.
The frequency of the hot resonator was taken to be $f_h=8.237$ GHz, i.e. it is two times higher than the frequency
of the diagnostic resonator of the Sample II. % (see the discussion at the end of Sec. \ref{experiment}). 
We have also increased the temperature difference between the left and the right resistors by choosing  $T_L=176$ mK and  $T_R=156$ mK.
Such increase in $T_L-T_R$ is within the experimental uncertainity.
Finally, we have reduced the Josephson energy to $E_J/h=3.63$ GHz to achieve better agreement between spectroscopy plots of
Figs. \ref{spectr} (experiment) and \ref{Power}c (theory). 
Photonic heat powers $P_L^{\rm ph}(\Phi)$ and $P_R^{\rm ph}(\Phi)$, obtained in this way, are plotted in Fig. \ref{Power}b.
They indeed behave similarly to the experimental ones, namely, they oscillate in opposite directions and the
average power $(P_L^{\rm ph}+P_R^{\rm ph})/2$ is almost independent of the flux. 
Theory modelling clarifies the origin of this effect. Indeed, according to Eq. (\ref{Pi})
the photonic heat powers dissipated in the resistors are expressed as $P_L^{\rm ph} = P_{Lh}^{\rm ph}+P_{LR}^{\rm ph}$, 
$P_R^{\rm ph}=P_{Rh}^{\rm ph}-P_{LR}^{\rm ph}$.
Numerically we find that the currents flowing from the hot resistor to the left and the right ones, $P_{Lh}^{\rm ph}$ and $P_{Rh}^{\rm ph}$, 
weakly depend on magnetic flux because of the strong
detuning between the qubit and the hot resonator. Thus, the  flux dependence of the powers $P_L^{\rm ph}$ and $P_R^{\rm ph}$ 
predominantly comes from the photonic heat current flowing from the right to the left resistor $P_{LR}^{\rm ph}(\Phi)$, which contributes
to them with opposite signs due to energy conservation. Comparing  Figs. \ref{Power}a and b, we notice the similarity in the shape
of the experimental and the theoretical power-flux dependences for the right resistor.
However, the theoretical curve for the left resistor is inverted and shifted by one half of the flux quantum relative to
the experimental one. 
By tuning the system parameters further one can, in principle, reproduce the right shape of the $P_L^{\rm ph}(\Phi)$ dependence.

In Fig. \ref{Power}c we plot the sum of the two transmission probibilities $\tau_{Lh}(\omega,\phi)+\tau_{Rh}(\omega,\Phi)$ (\ref{tau})
evaluated with the same fit parameters as in the theory plots of Fig. \ref{Power}b.
We note close resemblence of this plot with Fig. \ref{spectr}, in which the measured value of the transmission coefficient $|S_{21}(\omega,\Phi)|$
is presented. This similarity is expected because the two values are approximately related as $|S_{21}|^2 \approx 1 - \alpha(\tau_{Lh}+\tau_{Rh})$,
where the constant $\alpha$ depends on the system parameters. It also
demonstrates that our model rather accurately describes the system even though it neglects
the anharmonicity of the qubit.
In addition, Fig. \ref{Power}c helps to clarify the origin of the unusual non-sinusoidal shape of the power oscillations in Figs. \ref{Power}a and b.
Indeed, since the qubit and the resonator frequencies do not cross each other and the contribution
of the qubit spectral lines to the heat transport is quite weak, 
one can rather accurately expand the heat current $P_{LR}^{\rm ph}$ in powers of 
the resonance frequency $\omega_0(\Phi)$, 
$P_{LR}(\Phi)\approx P_{LR}^{(0)} + A\omega_0^2(\Phi)$, where $A$ is a pre-factor. 
The last term in this expression $\propto \omega_0^2(\Phi)$ produces the cusps visible 
in $P_{L,R}(\Phi)$ dependencies both in Figs. \ref{Power}a and b. They occur at $\Phi=\Phi_0/2$
where the dependence $\omega_0(\Phi)$ also has a cusp, see the lowest yellow line in Fig. \ref{Power}c.
The cusps become a bit rounded if one proceeds more accurately and 
replaces the frequency $\omega_0(\Phi)/2\pi$ by the qubit tranistion frequency $f_{01}(\Phi)$
in the expansion.% As we discussed in Sec. \ref{spectroscopy} 
At the flux point $\Phi=0.5\Phi_0$ the harmonic
approximation becomes inaccurate. Indeed, at this point the qubit frequency $\omega_J/2\pi$ is a bit higher than 
the frequency $f_{01}$ computed with full quantum approach, cf.  Figs. \ref{spectr} and \ref{Power}c.

The absolute values of experimentally measured (Fig. \ref{Power}a) powers exceed the  numerically estimated ones (Fig. \ref{Power}b) by
a factor $\sim 10^3$. Thus, most of the heat power between the resistors is transmitted by the substrate phonons or by other mechanisms
not included in our model. 
It is typical for this type of experiments, see e.g. Refs. \cite{QHV,QHR}.
Indeed, one has to heat the system strongly to make the small photonic heat flux measurable, 
but other contributions to the heat flux grow even stronger with rising temperature.
The theoretical power modulation amplitudes also differ the experimental ones: we find $\delta P_L\approx 13$ aW  
and $\delta P_R\approx 10$ aW in the experiment,
and $\delta P_L^{\rm ph}=\delta P_R^{\rm ph}\approx 0.54$ aW in the simulation. 
In theory one can increase the power modulation by, for example, increasing the qubit frequency so that it crosses
the frequencies of the resonators. 
In general, we have noticed that the model systematically underestimates the modulation amplitude of the powers $P_L$ and $P_R$.
Further research is required in order to resolve this problem.

\section{Discussion}
\label{Summary}

We have studied the heat transport by photons in a three terminal system containing a flux qubit
realized as superconducting loop with three identical Josephson junctions.
We have combined the DC heat transport measurements with the microwave spectroscopy performed on
a separate sample with nominally identical parameters. In this way, we have verified that the
flux qubit operates as a quantum system with its level spacing modulated by magnetic flux, and related
this effect to the modulation of the heat power in DC measurements. 
Employing the standard theory models, we have described both the changes in the qubit
spectrum and the modulation of the heat flux by magnetic field with the same set of parameters.  
Our experiment is an important step towards practical realization of
on-chip quantum heat transistors, thermal amplifiers and heat pumped masers.

\section{Acknowledgement}
We acknowledge valuable discussions with J. Ankerhold and G. Kurizki. 
This work was supported by Marie Sklodowska-Curie grant agreement No 843706, by the Academy of Finland Centre of Excellence
program (project 312057), and by the European Union's Horizon 2020 research 
and innovation programme under the European Research Council (ERC) programme (grant agreement 742559).

\section{Methods}
\textbf{Fabrication}.
The devices are fabricated using three steps of electron-beam lithography (EBL) to create a mask on a silicon substrate. First EBL mask to prepare the niobium pattern and the groundplane of the device: feedline, resonators, cross-shape island and superconducting probes for thermometry are formed by reactive ion etching a 200 nm niobium metal film deposited by DC magnetron sputtering onto the high-resistivity silicon wafer. A 20-nm-thick aluminium film has been grown by atomic layer deposition on the wafer prior to Nb sputtering. The characteristic impedance of the coplanar waveguides used for the feedline and resonators is 50 $\Omega$. Second, the SQUID with three Josephson junctions is realized with shadow-mask EBL, formed using two layers of poly(methylmetalcrylate–methacrylic) acid P(MMA–MAA) resist spun for 60 s at 4000 rpm followed by one layer of polymethyl-metacrylate (PMMA) spun for 60 s at 4000 rpm, all resist layers are immediately baked at 160 C, and followed by thin film physical vapor deposition in an electron-beam evaporator with an intermediate oxidation, using the Dolan bridge technique. The evaporation was preceded by argon ion plasma milling to facilitate the clean contact between aluminium and niobium. Third, the normal metal resistors and the normal–insulator–superconductor (NIS) tunnel junction elements for thermometry are also patterned by EBL, using the same resist, and then similarly deposited onto the wafer in three steps: Al with an in situ oxidation to make NIS junctions, Cu for reservoirs and Al. The fabrication is completed by spin-coating a protective layer of photoresist (AZ5214E) and dicing by diamond-embedded resin blade. The resist was then removed using  acetone. The resistance of the SIS $\mathcal{R}=28 \pm 1 \,\mathrm{k\Omega}$ and NIS tunnel junctions $\mathcal{R}=12 \pm 1 \,\mathrm{k\Omega}$ was measured on the test structures of the same dimensions on the same chip as shown in Fig.  ~\ref{device}a. Both devices for heat transport measurements ( Fig.  ~\ref{device}a ) and RF characterization ( Fig.  ~\ref{device}b ) were fabricated simultaneously on the same wafer. 
  
\textbf{RF spectroscopy}.
The sample for RF spectroscopy was cooled in a BlueFors dilution refrigerator with a base temperature of 10 mK. In this design, a 7.4 GHz spectroscopy resonator connects the top terminal of the cross-shaped island to a feedline for spectroscopy characterization via transmission microwave readout. The spectroscopy measurements were performed using a vector network analyzer (VNA) at room temperature with the signal reaching the sample via an RF line and attenuated at different temperatures inside the cryostat. This scheme is presented in Fig. ~\ref{device}e. The output signal is then passed through two circulators at base temperature to a low noise HEMT amplifier mounted at 4 K, providing 46 dB gain, followed by an additional 28 dB amplifier outside of the cryostat. Characterization started by measuring the transmission $S_{21}$ through the feedline to identify the $\lambda/4$ diagnostic resonator. To characterize the qubit and qubit-resonator couplings we perform two-tone spectroscopy by first applying a microwave low-power probe tone, followed by a second tone, whose frequency is swept. The Josephson energy is tuned by applying the magnetic flux with an external coil.

\textbf{Thermal conductance measurements} were performed in a custom-made plastic dilution refrigerator with a mixing chamber (MXC) temperature varied in the range 90-400 mK.
The device is wire-bonded in a custom-made brass stage enclosed by two brass Faraday shields and fixed to the MXC with a proper thermalization. The readout scheme consists of thermocoax-filtered cryogenic lines with effective signal bandwidth of 0–10 kHz, for low-impedance loads. At room temperature, the voltage signal is amplified by  a low noise amplifier Femto DLVPA-100-F-D. Heating of the thermal reservoir is realized by DC / AC signals, which were applied by programmable function generators and read on the control thermal reservoirs by multimeter / lock-in amplifier, synchronised to the square-wave modulation $f \sim 77$ Hz of the heated voltage bias. 
Thermometry was performed in SINIS configuration \cite{revmodgiaz}, calibration of thermometers, presented in Fig. ~\ref{Calibration} (a,b,c), was done by monitoring the voltage while applying a current bias between the superconducting probes and varying the MXC temperature up to 400 mK. The device is well thermalized to the MXC, therefore  we can assume that the
phonon temperature is in equilibrium with the MXC temperature, which is monitored by a ruthenium oxide thermometer that has been calibrated against a Coulomb blockade thermometer.  The energy conservation among three thermal reservoirs is verified by performing three measurements: 1) heating the cold bath and performing local thermometry on the cold and remote on the drain and hot baths Fig. ~\ref{Calibration} (d,e,f); 2) heating and local thermometry on the drain bath and remote thermometry on the cold and hot baths Fig. ~\ref{Calibration} (g,h,i); 3) heating and local thermometry on the hot bath and remote thermometry on the cold and drain baths Fig. ~\ref{Calibration} (j,k,l). 
In Fig. ~\ref{Power} the magnetic flux is tuned by a superconducting solenoid encompassing the entire sample stage assembly, inside of a high-permeability magnetic shield, which is mounted inside the refrigerator at 4 K.

\section{Author contributions}
A.G., G.T., D.S.G., and J.P.P. conceived the experiment
and model and interpreted the results. 
A.G. designed, fabricated, and measured the samples with contributions from  J.T.P. and D.L..
Theoretical contributions were conceived and solved by D.S.G.  with inputs from G. T..
All authors have been involved in the analysis, and
discussion of results, and manuscript preparation.

\section{Competing financial interests}
The authors declare no competing financial interests.

\section{Data availability}
The data that support the plots within this article are available from the corresponding author upon reasonable request.


\begin{thebibliography}{}

\bibitem{Blais} Alexandre Blais, Arne L. Grimsmo, S. M. Girvin, Andreas Wallraff,
Circuit Quantum Electrodynamics,
\href{https://arxiv.org/abs/2005.12667}{arXiv preprint arXiv:2005.12667}
	
\bibitem{Meschke} M. Meschke, W. Guichard, and J. P. Pekola, Single-mode heat conduction by photons, 
\href{https://doi.org/10.1038/nature05276} {Nature {\bf 444}, 187 (2006).} 
% Broader scope theory..	
	
\bibitem{revmodgiaz} F. Giazotto, T. T. Heikkil\"a, A. Luukanen, A. M. Savin, J. P. Pekola, Opportunities for mesoscopics in thermometry and refrigeration: Physics and applications, \href{https://doi.org/10.1103/RevModPhys.78.217} {Rev. Mod. Phys {\bf 78}(1) 217 (2006).} 


\bibitem{Timofeev} A. V. Timofeev, M. Helle, M. Meschke, M. Möttönen, and J. P. Pekola, Electronic Refrigeration at the Quantum Limit, \href{https://doi.org/10.1103/PhysRevLett.102.200801} {Phys Rev. Lett. {\bf 102}, 200801 (2009).}  

\bibitem{Partanen} M. Partanen, K. Yen Tan, J. Govenius, R. E. Lake, M. K. Mäkelä, T. Tanttu, and M. Möttönen, Quantum-limited heat conduction over macroscopic distances, \href{https://doi.org/10.1038/nphys3642} {Nat. Phys. {\bf 12}, 460 (2016).}  


\bibitem{tan_quantum-circuit_2017} K. Y. Tan, \textit{et al.}, Nat. Comm. {\bf 8}, 15189 (2017).
%Engineering environments..

\bibitem{karimi_otto_2016} B. Karimi, J. P. Pekola,  Otto refrigerator based on a superconducting qubit: Classical and quantum performance, \href{https://doi.org/10.1103/PhysRevB.94.184503} {Phys. Rev. B {\bf 94}, 184503 (2016).}  
% Broader scope theory  


\bibitem{kosloff} R. Kosloff, A. Levy, Annual Rev. Phys. Chem. {\bf 65}, 365 (2014).
%Theory proposal..		

\bibitem{GiazottoRect}  A. Fornieri, M.J. Martínez-Pérez, F. Giazotto, AIP Advances {\bf 5}, 053301 (2015).
% Theory proposal..	

\bibitem{martinezheat}  M. J. Mart\'inez-P\'erez, F. Giazotto, Nature {\bf 492}, 401 (2012).
%caloritronics ..	

\bibitem{QHV} A. Ronzani, B. Karimi, J. Senior, Y.-C. Chang, J.T. Peltonen, C. Chen and J.P. Pekola, Tunable photonic heat transport in a quantum heat valve, \href{https://doi.org/10.1038/s41567-018-0199-4} {Nat. Phys. {\bf 14}, 991 (2018).}  

\bibitem{QHR} J. Senior, A. Gubaydullin, B. Karimi, J. T. Peltonen, J. Ankerhold, and J. P. Pekola, Heat rectification via a superconducting artificial atom, \href{https://www.nature.com/articles/s42005-020-0307-5}{Comm. Phys. {\bf 3}, 40 (2020).}
% Precursor experiment..	

\bibitem{Sanchez} Israel Diaz, Rafael Sanchez, The qutrit as a heat diode and circulator, arXiv:2109.06551. \href{https://arxiv.org/abs/2109.06551}{arXiv preprint arXiv:2109.06551}

\bibitem{Joulain} K. Joulain, J. Drevillon, Y. Ezzahri, and J. Ordonez-Miranda, 
Quantum Thermal Transistor,
\href{https://journals.aps.org/prl/pdf/10.1103/PhysRevLett.116.200601}{Phys. Rev. Lett. {\bf 116}, 200601 (2016).}

\bibitem{Zhang} Y. Zhang, Z. Yang, X. Zhang, B. Lin, G. Lin and J. Chen,
Coulomb-coupled quantum-dot thermal transistors,
\href{https://iopscience.iop.org/article/10.1209/0295-5075/122/17002}{Europhys. Lett. {\bf 122}, 17002 (2018).}


\bibitem{Majland} M. Majland, K. S. Christensen,  and N.T. Zinner, Quantum thermal transistor in superconducting circuits, \href{https://journals.aps.org/prb/abstract/10.1103/PhysRevB.101.184510}{Phys. Rev. B {\bf 101}, 184510 (2020).}

\bibitem{KurizkiPRR} M.T. Naseem, A. Misra, M. Avijit, O. M\"ustecaplio\ifmmode \breve{g}\else \u{g}\fi{}lu, G. Kurizki, Minimal quantum heat manager boosted by bath spectral filtering,\href{https://journals.aps.org/prresearch/abstract/10.1103/PhysRevResearch.2.033285} {Phys. Rev. Research {\bf 12}, 033285 (2020)}.

\bibitem{Liu} Huan Liu, Chen Wang, Lu-Qing Wang, and Jie Ren, 
Strong system-bath coupling induces negative differential thermal conductance and heat amplification in nonequilibrium two-qubit systems,
\href{https://journals.aps.org/pre/abstract/10.1103/PhysRevE.99.032114}{Phys. Rev. E {\bf 99}, 032114 (2019).}

\bibitem{Wang} Chen Wang, Dazhi Xu, Huan Liu, and Xianlong Gao, 
Thermal rectification and heat amplification in a nonequilibrium V-type three-level system,
\href{https://journals.aps.org/pre/abstract/10.1103/PhysRevE.99.042102}{Phys. Rev. E {\bf 99}, 042102 (2019).}

\bibitem{kosloff2} A. Levy and R. Kosloff, Quantum Absorption Refrigerator,
\href{https://journals.aps.org/prl/abstract/10.1103/PhysRevLett.108.070604}{Phys. Rev. Lett. {\bf 108}, 070604 (2012).}

\bibitem{Segal} M. Kilgour and D. Segal, Coherence and decoherence in quantum absorption refrigerators,
\href{https://journals.aps.org/pre/abstract/10.1103/PhysRevE.98.012117}{Phys. Rev. E {\bf 98}, 012117 (2018).}

\bibitem{Scovil1959} H. E. D. Scovil and E. O. Schulz-DuBois, Three-Level Masers as Heat Engines, \href{https://link.aps.org/doi/10.1103/PhysRevLett.2.262}{Phys. Rev. Lett. {\bf 2}, 262 (1959).}	

\bibitem{George} G. Thomas, A. Gubaydullin, D.S. Golubev, and J.P. Pekola, Thermally pumped on-chip maser,
\href{https://link.aps.org/doi/10.1103/PhysRevB.102.104503} { Phys. Rev. B {\bf 102}, 104503 (2020).}	

\bibitem{Orlando} T.P. Orlando \textit{et al.}, Superconducting persistent-current qubit, 
\href{https://doi.org/10.1103/PhysRevB.60.15398} {Phys. Rev. B {\bf 60}, 15398–15413 (1999).}

\bibitem{Esteve2008}  A. Palacios-Laloy, F. Nguyen, F. Mallet, P. Bertet, D. Vion and D. Esteve, 
%Tunable Resonators for Quantum Circuits
\href{ https://doi.org/10.1007/s10909-008-9774-x}{J. Low Temp. Phys.  {\bf 151} 1034 (2008)}.
  
	
\bibitem{Barends} R. Barends, J. Kelly, A. Megrant, D. Sank, E. Jeffrey, Y. Chen, Y. Yin, B. Chiaro, J. Mutus, C. Neill, P. O’Malley, P. Roushan, J. Wenner, T. C. White, A. N. Cleland, John M. Martinis, 
\href{https://doi.org/10.1103/PhysRevLett.111.080502}{Phys. Rev. Lett. {\bf 111}, 080502 (2013).}	
	
\bibitem{Pascal}  L. M. A. Pascal, H. Courtois, and F. W. J. Hekking, 
Circuit approach to photonic heat transport,
\href{https://journals.aps.org/prb/abstract/10.1103/PhysRevB.83.125113}{Phys. Rev. B {\bf 83}, 125113 (2011).}

\bibitem{Thomas} G. Thomas, J. P. Pekola, and D. S. Golubev,
Photonic heat transport across a Josephson junction,
\href{https://journals.aps.org/prb/abstract/10.1103/PhysRevB.100.094508}{Phys. Rev. B {\bf 100}, 094508 (2019).}

\bibitem{Koch} J. Koch, T. M. Yu, J. M. Gambetta, A. A. Houck, D. I. Schuster, J. Majer, A. Blais, M. H. Devoret, S. M. Girvin, and R. J. Schoelkopf, Charge-insensitive qubit design derived from the Cooper pair box, \href{https://journals.aps.org/pra/abstract/10.1103/PhysRevA.76.042319}  { Phys. Rev. A {\bf 76}, 042319 (2007).}

\bibitem{Probst2015} S. Probst, F. B. Song, P. A. Bushev, A. V. Ustinov, and M. Weides,  Rev Sci Instrum {\bf 86}, 024706 (2015).


\end{thebibliography}
\end{document}